\documentclass[aps,showpacs,superscriptaddress,nofootinbib,a4paper,twocolumn,10pt]{revtex4}

\usepackage{amsmath}
\usepackage{graphicx}

\begin{document}

\title{Semiclassical statistico-dynamical description of polyatomic
 photo-dissociations: State-resolved distributions.}

\author{Maykel Leonardo \surname{Gonz\'alez--Mart\'{\i}nez}}
\email[Corresponding author, ]{E-mail: mleo@instec.cu}
\affiliation{Departamento de F\'{\i}sica General, Instituto Superior de
 Tecnolog\'{\i}as y Ciencias Aplicadas, Habana 6163, Cuba}
\affiliation{Institut des Sciences Mol\'eculaires, Universit\'e Bordeaux 1, 351
 Cours de la Lib\'eration, 33405 Talence Cedex, France}

\author{Laurent Bonnet}
\email[Corresponding author, ]{E-mail: l.bonnet@ism.u-bordeaux1.fr}
\affiliation{Institut des Sciences Mol\'eculaires, Universit\'e Bordeaux 1, 351
 Cours de la Lib\'eration, 33405 Talence Cedex, France}

\author{Pascal Larr\'egaray}
\affiliation{Institut des Sciences Mol\'eculaires, Universit\'e Bordeaux 1, 351
 Cours de la Lib\'eration, 33405 Talence Cedex, France}

\author{Jean--Claude\ Rayez}
\affiliation{Institut des Sciences Mol\'eculaires, Universit\'e Bordeaux 1, 351
 Cours de la Lib\'eration, 33405 Talence Cedex, France}

\date{\today}

\begin{abstract}
\noindent An alternative methodology to investigate indirect polyatomic
processes with quasi-classical trajectories is proposed, which effectively
avoids any binning or weighting procedure while provides rovibrational
resolution.  Initial classical states are started in terms of angle-action
variables to closely match the quantum experimental conditions and later
transformed into Cartesian coordinates, following an algorithm very recently
published [\emph{J.\ Chem.\ Phys.}~\textbf{130}, 114103 (2009)].  Trajectories
are then propagated using the `association' picture, \emph{i.e.}~an inverse
dynamics simulation in the spirit of the exit-channel corrected phase space
theory of Hamilton and Brumer [\emph{J.\ Chem.\ Phys.}~\textbf{82}, 595 (1985)],
which is shown to be particularly convenient.  Finally, an approximate
quasi-classical formula is provided which under general conditions can be used
to add possible rotational structures into the vibrationally-resolved
quasi-classical distributions.  To introduce the method and illustrate its
capabilities, correlated translational energy distributions from recent
experiments in the photo-dissociation of ketene at 308~nm
[\emph{J.\ Chem.\ Phys.}~\textbf{124}, 014303 (2006)] are investigated.  Quite
generally, the overall theoretical algorithm reduces the total number of
trajectories to integrate and allows for fully theoretical predictions of
experiments on polyatomics.
\end{abstract}


\maketitle

\section{Introduction}
\label{sec:introduction}
The 20th century witnessed both the advent of modern computers and the
associated exponential growth in computational power, which continues nowadays.
As more can be done in significantly less time, simulation of inherently more
complex processes and/or larger systems is attempted.  In the field of chemical
physics, what started as describing atomic and diatomic molecular processes
currently looks after the simulation of, \emph{e.g.}~larger polyatomics,
clusters, phenomena at surfaces and biomolecular complexes of interest.

However, despite all the effort being devoted to develop approaches which
include quantum effects in the dynamics of large molecular systems
\cite{whmiller:06}, realistic quantum-mechanical \emph{simulations} of
polyatomic processes remain prohibitively expensive.  To date, the only general
theoretical alternative is the quasi-classical trajectory method (QCT)
\cite{rnporter:76,dgtruhlar:79,tdsewell:97}, which is based on numerically
solving Hamilton's (or Newton's) equations of motion.  Primarily, this is due to
the cumbersome basis sets that need to be handled within quantum calculations in
order to avoid crude approximations, but also partially because developing
general, \emph{i.e.}~\emph{scalable} classical trajectory codes is relatively
easier.

To allow for comparison with experiments, both the selection of classical
initial conditions and the statistical treatment of QCT results should be
semiclassical in spirit.  In particular, to mimic quantum features using
classical (continuous) magnitudes, additional \emph{binning} or \emph{weighting}
procedures are quite difficult to avoid.  As implied, the underlying source of
the problem---as with some other drawbacks of QCT---is that classically,
vibrations and rotations are not quantised \emph{per se}.  In this regard,
important advances have been made in the last few years by replacing the
standard/histogram binning (SB) procedure by a Gaussian weighting (GW)
\cite{lbonnet:97,lbannares:03,lbonnet:04a,txie:05,mleo:07b,lbonnet:08,mleo:08,zsun:08}.
In the latter, the different trajectories are assigned different (Gaussian)
statistical weights, larger the closer its final actions to integer values.
Despite its many promising capabilities, we have recently pointed out how the
convergence of GW could be affected when applied to processes involving
polyatomics \cite{mleo:08}.  This is mainly because increasing the number of
atoms in a molecule will almost certainly increase the number of modes for which
a quantum--mechanical description is convenient---and Gaussian weights must be
assigned.  As a result, the overall weight of a given trajectory could be
\emph{significantly} reduced, in particular if several actions vary
simultaneously.

At the same time, powerful experimental techniques have been developed during
the precedent decades and very precise measurements of state-to-state
observables are now possible.  Correlated product distributions---where final
state and/or translational energy distributions are measured in coincidence with
given rovibrational states of the products---are among such.  These
increasingly sophisticated experiments demand finer statistics on the QCT
results for highly resolved, intrinsically quantum features are much more
difficult to `converge' using classical mechanics.  The `brute force' approach
is of course to increase the number of trajectories in the classical
ensemble---and as their propagation is easily task-parallelised,
\emph{i.e.}~particularly suitable to solve in computer clusters---running
millions of such is becoming common.

In this work we report on an alternative QCT algorithm for investigating
polyatomic processes which effectively avoids any binning or weighting method
while provides both vibrational and rotational resolution.  Its main advantage
comes from the fact that initial conditions are directly generated at the
semiclassical rovibrational states of interest in terms of angle-action
variables, which automatically reduces the total number of trajectories to
integrate.  These are however propagated in Cartesian coordinates, by means of a
transformation very recently published \cite{mleo:09a}.  In addition, a
quasi-classical formula is derived to further incorporate possible rotational
structures.  To introduce and illustrate the capabilities of our method, we
simulate highly-structured empirical correlated product distributions from the
dissociation of ketene at 308~nm \cite{avkomissarov:06}.

The paper is organized as follows: the problem, theoretical aspects of the QCT
implementation and results of its application are dealt with in
Section~\ref{sec:theory} and later discussed in Section~\ref{sec:discussion}; at
last, Section~\ref{sec:summary} summarizes.

\section{THEORY AND APPLICATION}
\label{sec:theory}
\subsection{Ketene dissociation at the $S_0$ electronic surface}
\label{sec2:system}
The photo-fragmentation of ketene,
\begin{eqnarray*}
\mathrm{CH}_2\mathrm{CO}(\tilde{X}{}^1A_1)
                         &\stackrel{h\nu}{\longrightarrow}&
\mathrm{CH}_2\mathrm{CO}(\tilde{A}{}^1A'')\\
                         &\stackrel{\mathrm{IC}}{\longrightarrow}&
\left\{\begin{array}{l} \mathrm{CH}_2(\tilde{a}{}^1A_1)+
                                                      \mathrm{CO}(X^1\Sigma_1)\\
   \mathrm{CH}_2(\tilde{X}{}^3B_1)+\mathrm{CO}(X{}^1\Sigma_1) \end{array}\right.
\end{eqnarray*}
is arguably the most intensively investigated polyatomic process in the last
decades,
\emph{cf.}~\cite{avkomissarov:06,vzabransky:75,dnesbitt:85,hbitto:86,i-cchen:88,whgreen:88,edpotter:89,sjklippenstein:89,skkim:90,whgreen:90,sjklippenstein:90,igarcia-moreno:94a,igarcia-moreno:94b,cgmorgan:96a,cgmorgan:96b,sjklippenstein:96,eawade:97a,eawade:97b,mlcosten:00,kmforsythe:01,jliu:05}.
At present, it is firmly established that following optical excitation to the
$\tilde{A}{}^1A''$ state, the molecule either undergoes intersystem crossing to
the lowest-lying, triplet, or fast internal conversion to the singlet electronic
state.  From these, dissociation into methylene and carbon monoxide occurs.
Despite the triplet threshold lies $\sim$3150 cm$^{-1}$ below the singlet, the
fact that it presents a (small) barrier to dissociation---of a few cents of
inverse centimetres---, but mainly, that the interaction which accounts for the
change in electronic spin is rather weak, makes the singlet channel
statistically dominant from excess energies as low as 100--200~cm$^{-1}$.  Such
conditions make this process an effective prototype for a
\emph{barrierless polyatomic} unimolecular reaction on a \emph{single} potential
energy surface (PES).  Experimentally, it has the additional advantage of a
well-defined total energy and time origin distinctive of photo-dissociations
\cite{avkomissarov:06}.

Earlier measurements of energy-dependent rates and overall product distributions
were used as benchmarks to test and develop various statistical models including
phase space theory (PST) and Rice--Ramsperger--Kassel--Marcus (RRKM) theory,
\emph{cf.}~\cite{sjklippenstein:89,skkim:90,whgreen:90,sjklippenstein:90}.  It
was concluded that under moderate excitations the molecule dissociates
statistically, exit-channel effects becoming increasingly important with
augmenting energy.  The dissociating picture therefore evolves from that of a
loose to a tight transition state (TS) \cite{eawade:97b}.  During the last
decade, however, there has been growing interest in the problem of correlated
product distributions \cite{cgmorgan:96a,mlcosten:00,avkomissarov:06}.  In the
most recent experiments, Komissarov \emph{et al.}~\cite{avkomissarov:06} studied
the photo-fragmentation of the molecule at 308~nm and measured product
translational energy distributions for specific rotational states of CO,
\emph{i.e.}~$P(E_\mathrm{t};j_\mathrm{CO})$.  These are the observables we will
focus on in the present work.

\subsection{Singlet potential energy surface}
\label{sec2:pes}
Extensive trajectory calculations require a reliable numerical representation of
the singlet PES.  This was generated from the most recently reported high-level
\emph{ab initio} data \cite{sjklippenstein:96}.  In this work, Klippenstein,
East and Allen used a separation in transitional and conserved modes to
determine the quadratic force fields along the reaction path---at the
CCSD/[5s4p2d, 4s2p] level of theory---in terms of $C^I_s$ symmetry internal
coordinates.  For excess energies of the order of that obtained with a 308~nm
laser ($E_\mathrm{exc}\approx2350$~cm$^{-1}$), these force fields should provide
all essential information on the couplings between the intra-fragment modes and
the dissociation coordinate in the vicinity of the reaction path, as well as the
interaction dependence on the mutual orientation of the fragments\footnote{We
discuss the implications of this quadratic approximation to the PES for
trajectory calculations in some detail in Section~\ref{sec3:lambdaPES}.}.  To
further assess its quality, in the same work, the authors showed that
variational RRKM energy-dependent rates predicted with this PES are in
quantitative agreement with the experiment.

Being $\boldsymbol{S}=\{S_i\}$ the symmetry internal coordinates, $F_{ij}$
($i,j=\overline{1,9}$) the quadratic force constants and $V_\mathrm{CC}$
(Fig.~7 in Ref.~\cite{sjklippenstein:96}) the minimum potential energy along the
reaction path as a function of the C--C interatomic distance, $S_4$, the PES may
be written as
\begin{equation}
 V(\boldsymbol{S})=V_\mathrm{CC}(S_4)+\frac{1}{2}\sum^{9}_{i=1}
            \sum^{9}_{\substack{j=1\\ j\neq 4}}F_{ij}(S_i-S_{0i})(S_j-S_{0j}),
\end{equation}
where most $S_{0k}=S_{0k}(S_4)$ can be straightforwardly calculated from the
definitions of symmetry internal coordinates and the analytic parametrizations
for the reaction path (Tables VI and III, \emph{ibid.}).  The slightly more
elaborate are
$S_{06}=\sqrt{2}\arccos{\left[-\cos{(\alpha/2)}\cos{\gamma}\right]}$ and
$S_{09}=\sqrt{2}
           \arccos{\left[-\cos{(\alpha/2)}\sin{\gamma}\sin^{-1}{\beta}\right]}$,
where---if individual ketene atoms are numbered as they enter in the chemical
formula---$\alpha=\angle($H$^1$C$^1$H$^2)$,
$\beta=\beta_{1,2}=\angle($H$^{1,2}$C$^1$C$^2)$ and
$\gamma$ (CH$_2$ wagging angle) are all angular internal coordinates defined in
Fig.~1 of the same work.  The $F_{ij}$ functions were obtained from fittings to
values in Table VI, \emph{ibid.}

Finally, we validated the PES by diagonalising the Hessian matrix and confirming
that all local harmonic frequencies previously reported (Table VII in
Ref.~\cite{sjklippenstein:96}) were successfully reproduced.

\subsection{Quasi-classical trajectories}
\label{sec2:qct}
We will introduce in what follows a particular implementation of QCT ideas and
show its predictive power in the simulation of highly-resolved
$j_\mathrm{CO}$-correlated translational energy distributions, from the
unimolecular dissociation of ketene.  Theoretically, it can be assumed that
under the experimental conditions reported in Ref.~\cite{avkomissarov:06}, these
product distributions correspond to a given total energy
$E=E_\mathrm{exc}+E_\mathrm{nmodes}$ and angular momenta $J$,
\emph{i.e.}~$P(E_\mathrm{t};j_\mathrm{CO})\equiv P(E_\mathrm{t};E,J,j_\mathrm{CO})$.
The zero-point energy in the molecule normal modes,
$E_\mathrm{nmodes}\approx4730$~cm$^{-1}$.

In reading the following section, it is important to keep in mind that we use an
association rather than a dissociation picture for propagating trajectories.
This is based on a method due to Hamilton and Brumer \cite{ihamilton:85} which
will be introduced in Section~\ref{sec3:eccpst}, where its many advantages are
also discussed.

\subsubsection{Initial conditions: Angle-action variables and the transformation
 to Cartesian coordinates}
\label{sec3:aa2cart}
Within QCT, it is always preferred to generate the ensemble of classical initial
conditions for the very precise quantum states explored in the experiment.  Such
a generation is \emph{naturally} performed in terms of angle-action variables
\cite{whmiller:74,hgoldstein:80,dmwardlaw:85,mschild:91}, the actions being the
classical counterpart of quantum numbers\footnote{Throughout this work all
actions will be quoted in $\hbar$ units.}.  These variables, however, should
preferably not be used to run trajectories for they may lead to strong numerical
instabilities.  In fact, Cartesian coordinates are ideal to solve the classical
equations of motion for molecular processes involving polyatomics.  Codes built
on these coordinates should result easily scalable and standard numerical
approaches, adequate.  Therefore, the transformation from angle-action variables
to Cartesian coordinates is an important step in QCT.  Very recently, we
reported on such a transformation \cite{mleo:09a} that fits to our purposes
here.

Discarding all center-of-mass coordinates the phase-space results
\begin{eqnarray}
\label{eq:gamma5at}
 \mathbf{\Gamma}&=&
 \{R,q_i,\alpha,\beta,\alpha_l,\alpha_1,\alpha_2,\alpha_k,\gamma_1\} \nonumber\\
 && \otimes\{P,x_i,J,J_z,l,j_1,j_2,k,\kappa_1\};\; i=\overline{1,4},
\end{eqnarray}
where $R$ and $P$ are the dissociation coordinate---the CH$_2$ and CO fragments
centres-of-mass relative distance---and conjugate momentum.  The remaining 22
angle-action variables are \cite{mleo:09a}: $q_i$, $x_i$, the vibrational phase
and corresponding action of the $i$th normal mode of CH$_2$ ($i=\overline{1,3}$)
and CO ($i=4$); $J$, $J_z$, the moduli of the total angular momentum and its
$z$-axis projection in the laboratory frame; $l$, the modulus of the orbital
angular momentum; $j_1$, $j_2$, the moduli of CH$_2$ and CO rotational angular
momenta; $k$, the modulus of the total rotational angular momentum
$\boldsymbol{k}=\boldsymbol{j}_1+\boldsymbol{j}_2$; and $\kappa_1$, the
algebraic value of the projection of $\boldsymbol{j}_1$ onto the $C_2$ axis of
CH$_2$.  Finally, $\alpha$, $\beta$, $\alpha_l$, $\alpha_1$, $\alpha_2$,
$\alpha_k$ and $\gamma_1$ are respectively all the angles conjugate to the
angular momenta mentioned above.  It is apparent that in terms of these
variables we can effectively set up, individually, \emph{all} the vibrational
\emph{and} rotational modes of the fragments.

Given $E$, $J$ (fixed to 1) and $j_2$ ($\equiv j_\mathrm{CO}$)---for each set of
energetically permitted vibrational states $x_i$, $i=\overline{1,4}$---the total
available energy $E_\mathrm{avail}$ and maximum allowed values $l_\mathrm{max}$
and $j_{1,\mathrm{max}}$ are completely determined\footnote{The choice of $R$
will become apparent in Sections \ref{sec3:eccpst} and \ref{sec3:integration}.}.
From all these, $J_z$, $k$ and $\kappa_1$ are thus randomly selected from
uniform distributions according to $J_z\in[-J,J]$,
$|j_{1,\mathrm{max}}-j_2|\le k\le j_{1,\mathrm{max}}+j_2$ and
$\kappa_1\in[-j_{1,\mathrm{max}},j_{1,\mathrm{max}}]$.  Microcanonical
distributions are finally obtained by aleatory selection of all angles and
phases from a uniform distribution in $[0,2\pi]$.  All vector constraints on the
set of angular momenta must also be satisfied.

At this point, the $\mathbf{\Gamma}$ vector is transformed to Cartesian
coordinates, using the algorithm presented in Ref.~\cite{mleo:09a}.  Here, a
slight readjustment is necessary due to the conditions of the experiment, as
the \emph{total energy} of the system---rather than the relative velocity
between the fragments---must be \emph{fixed}.  Specifically, the three Cartesian
components of $\boldsymbol{P}$ (Section II C 3 in Ref.~\cite{mleo:09a}) should
be computed right before the final step (determination of nuclear momenta,
second part of Section II C 11, \emph{ibid.}) for only at this point it is
possible to calculate the energy available to translation, $E_\mathrm{avail,t}$,
from which $P=\sqrt{2\mu E_\mathrm{avail,t}}$ ($\mu$ being the system reduced
mass).  Computationally, an incubation period is needed before using an
otherwise acceptable initial state to avoid possible correlations in the
numerical generation of pseudo-random numbers.

\subsubsection{Propagation: Exit-channel corrected phase space theory}
\label{sec3:eccpst}
As previously seen (Section \ref{sec2:system}), it is possible to assume that
the photo-excited ketene forms a long-lived intermediate complex prior to
fragmentation, which justifies the use of a microcanonical distribution at the
TS.  In this work, extensive trajectory calculations were performed in
conformity to the so-called exit-channel corrected phase space theory (ECCPST),
proposed in the 1980s by Hamilton and Brumer \cite{ihamilton:85}.  Within this
method, a microcanonical set of trajectories are initiated in the products and
directed towards the unimolecule region, the statistics being performed with
those `successful' trajectories reaching the TS.  Clearly, the underlying
assumption here is that the time-reversed dynamics of this subset of
trajectories yields a microcanonical distribution at the TS \cite{ihamilton:85}.
The method bears certain similarities with the statistical quasi-classical
trajectory (SQCT) model for atom-diatom insertion reactions of Aoiz and
co-workers \cite{fjaoiz:07a}.

Besides its simplicity, the ECCPST offers quite a few practical advantages,
\emph{e.g.}~(1) initial conditions are generated where the assumptions for
partitioning into product normal modes are best fulfilled---\emph{i.e.}~at the
inter-fragment distance for which the interaction between the CH$_2$ and CO is
negligible---so the usual dispersion in the ensemble's total energy will be
minimum (of about 2--3~cm$^{-1}$ in our calculations); (2) all threshold
behaviour is expected to be reproduced at its best, since product states are
given precise energies in direct correspondence to their quantum counterparts;
(3) initial conditions are straightforwardly defined by fixing the
centres-of-mass distance between the resulting fragments---which explicitly
appears in the transformation from angle-action variables to Cartesian
coordinates \cite{mleo:09a}---while the TS can be re-defined at convenience
during propagation (in this case, by the C--C interatomic distance which was
proven best \cite{sjklippenstein:96}); (4) since product states statistics is
performed on a subset of initial conditions, the extra (non-trivial) work of
discretising classical magnitudes at the end of the trajectories is avoided; (5)
any binning or weighting to cope with the conceivable non-adiabaticity of
vibrations in the course of reaction---which was found to be of about 15--20\%
after some preliminary calculations---can be avoided; and (6) the PES does not
need to be determined within the reactant region.  A few other advantages will
be discussed in the following sections.

The main drawback, however, is that (possibly) many trajectories are propagated
which never actually reach the TS.  In fact, we found that approximately only 1
out of 10 trajectories enters the reactants region\footnote{The indetermination
on the exact TS location makes necessary to discard some trajectories even when
propagating outwards from the reactants region.  In this case, these constitute
nearly one half of the total ensemble.}.  The overall propagation time may thus
be significantly reduced by applying strategies to quickly identify those
trajectories which will ultimately be discarded.

\subsubsection{Computational details: integration and convergence}
\label{sec3:integration}
Except when stated contrary, batches of $5\times10^{4}$ trajectories were
started at $R^{\infty}=10$~\AA\ and propagated towards the unimolecule region in
accordance with ECCPST (Section~\ref{sec3:eccpst}).  Convergence of reported
product translational energy distributions was tested by confirming that those
obtained with $2\times10^{5}$ trajectories are statistically identical.

A fixed-step fourth-order Adams-Bashforth-Moulton predictor-corrector method
initialised with a fourth-order Runge-Kutta integrator was used for propagation.
Trajectories were followed until one of two conditions was fulfilled: (1) the
molecule reached the transition state (at the C--C interatomic distance
$R^\ddagger\equiv R_\mathrm{CC}$, as relevant to the specific excitation energy
from Table X in Ref.~\cite{sjklippenstein:96}); and (2)
$t=T_\mathrm{max}=20$~ps.  The step size $\Delta t=2\times10^{-4}$~ps was
adjusted so that the maximum error in total energy did not exceed
$\Delta E=20$~cm$^{-1}$, \emph{i.e.}~less than 0.3\% of $E$, with which the mean
error was $\left<\Delta E\right>=5$~cm$^{-1}$.  Under these conditions, the
fraction of non-fragmented to fragmented trajectories was consistently smaller
than 0.5\%.  Typically, a single trajectory required between 1 and 3 seconds of
CPU time---according to the specific values of $x_2$ and $j_2$---on an Altix
XE210 computer with two Dual-Core Intel\textsuperscript\textregistered\
Xeon\textsuperscript\textregistered\ 5030 (`Dempsey') processors.

\begin{figure}[!t]
\includegraphics[width=85mm]{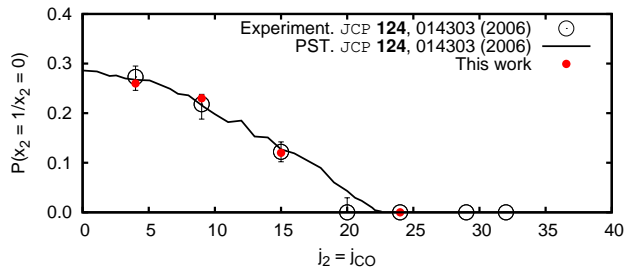}
\caption{(Colour online) Vibrational branching ratio for singlet CH$_2$ in
 coincidence with selected CO($x_4=0$, $j_2$) states.  Comparison with
 experimental and previous PST calculations.}
\label{fig:Px2_1_0}
\end{figure}

\begin{figure*}[!t]
\includegraphics[width=174mm]{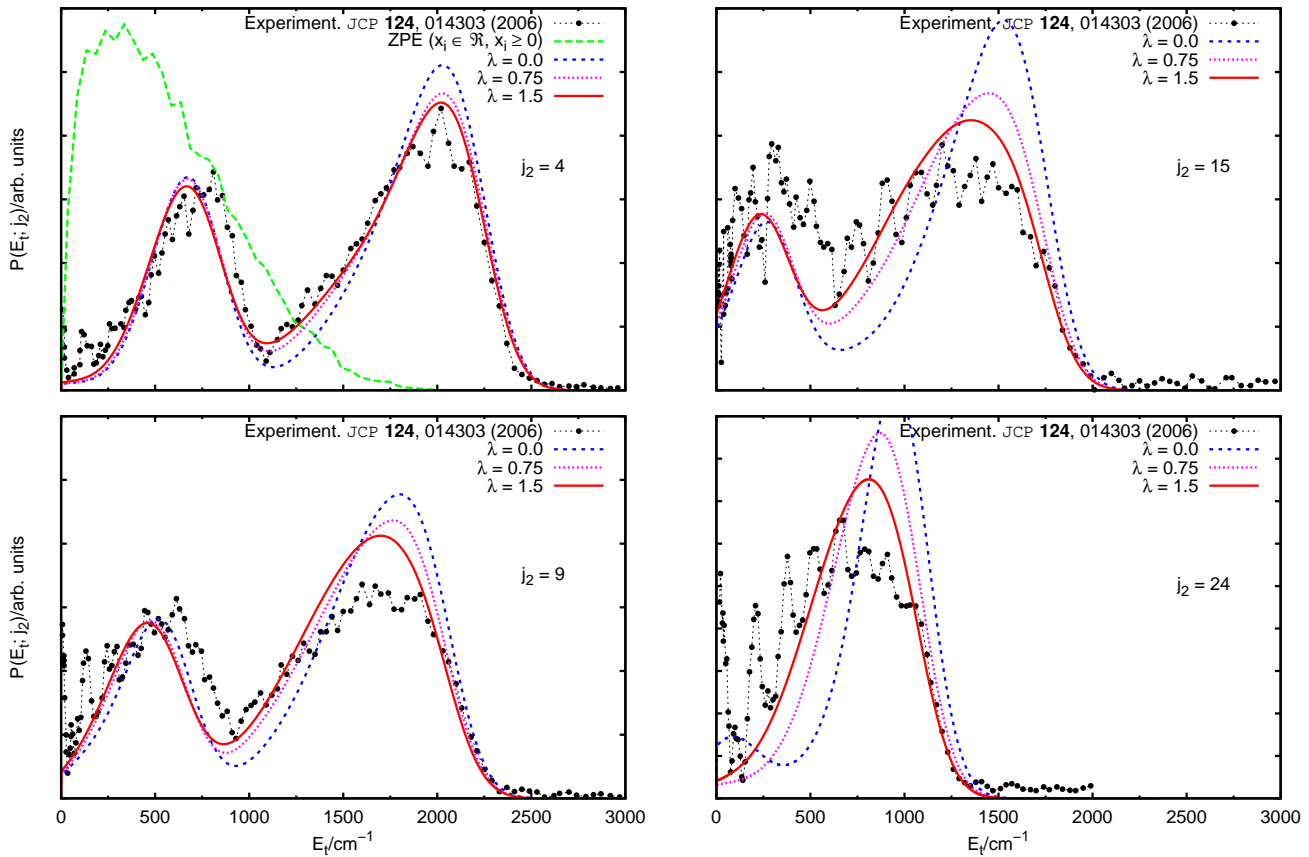}
\caption{(Colour online) Correlated product translational energy distributions
 after excitation with a 308~nm laser: QCT results with vibrational resolution.
 Comparison with the experiment.}
\label{fig:PEtrans_cnv}
\end{figure*}

\subsubsection{Correlated translational energy distributions}
\label{sec3:PEtrans}
After the subset of trajectories reaching the TS is identified, the
$P^C(E_\mathrm{t};E,J,x_2,j_\mathrm{CO})$ distribution corresponding to each
$x_2$ scissor mode of CH$_2$ is calculated by standard statistics over it.  From
these, the overall correlated translational energy distribution
$P^C(E_\mathrm{t};E,J,j_\mathrm{CO})$ is obtained by merging the different $x_2$
contributions altogether, with the corresponding weights according to the
vibrational branching ratios shown in Fig.~\ref{fig:Px2_1_0}.

In principle, to predict these weights at a given $j_2$ state of interest two
contributions need to be taken into account.  First, Monte Carlo methods should
be applied to find the relative probability that for each scissor state $x_2$ a
given set of angle-action variables is finally accepted as a valid initial
condition.  Here the ECCPST proves again particularly convenient, since this is
readily done if track is kept on the percent of successful trials and the
hypervolume in phase space from which each $x_2$ set of initial conditions was
randomly selected.  The second step is to correct this relative probability for
dynamical effects, which in practice would mean to determine the ratio of
capture probabilities from each scissor mode.  Despite its apparent simplicity,
obtaining a reliable approximation for the latter could be difficult for it not
only depends on the quality of the PES, but also on the relative accuracy of TS
locations and the set of integration parameters.  However, it was previously
shown that, by cancellation of errors, \emph{least-biased} values (also referred
as PST, depending on the convention employed) are in excellent agreement with
the experiment \cite{avkomissarov:06}.  The values reported in
Fig.~\ref{fig:Px2_1_0} are directly calculated from our initial conditions by
the algorithm explained at the top of this paragraph.  Their agreement with
experimental and previous theoretical values confirms the correctness of our
approach.  At last, the theoretical distributions are Gaussian-blurred to match
the experimental resolution.

Calculated curves for several representative $j_2$ states are compared with
measurements in Fig.~\ref{fig:PEtrans_cnv}---and have been quoted as
`$\lambda=0.0$' (blue), the reason for which will be given in the following
section.  The reported contributions from adjacent rotational states
\cite{avkomissarov:06} was appropriately taken into account to produce the
distributions corresponding to $j_2=4$ and 9.  Additionally, the
zero-point-energy (ZPE) corrected distribution \cite{ajcvarandas:94} has been
included in the $j_2=4$ panel, which illustrates the general behaviour for all
other rotational states.  To produce the latter, an ensemble of
$2\times10^{5}$ initial conditions with continuous actions was started in the
products, being the vibrational modes constraint between zero and their maximum
energetically allowed values.

\subsubsection{A modification to the PES}
\label{sec3:lambdaPES}
Except of course for the $j_2=24$ rotational state, all `$\lambda=0.0$'
theoretical curves in Fig.~\ref{fig:PEtrans_cnv} have the correct bimodal shape
due to the excitation of the first and the ground states of methylene's scissor
mode (with frequency $w_2\approx1400$~cm$^{-1}$).  The experimental positions of
the peaks are also fairly well reproduced.  It is important to stress here that
neither of these facts is trivial.  Note, for example, the strong discrepancy of
the ZPE-corrected curve which fails to reproduce any of them.  Moreover, to
achieve a similar agreement with the experiment in the case of the triatomic
dissociation NCO~$\longrightarrow$~{N + CO}, the GM method had to be applied
\cite{mleo:07b}.  However, the divergence between the experimental and ZPE
distributions has an additional important implication.  As the population
corresponding to the ground state of the CH$_2$ scissor mode $x_2=0$ is
practically negligible, even by using the GW one could never obtain the proper
translational energy distributions from ZPE-corrected calculations only.  It is
thus crucial that, at this point, the `vibrational' resolution has been
naturally obtained within QCT and no binning or weighting procedure is needed.

However, it is clear that within each vibrational lobe, these theoretical
distributions tend to predict larger than observed populations of low CH$_2$
rotational levels.  A plausible explanation may be derived from calculations on
triatomic complexes.  From these, it is known that quadratic representations of
PES systematically tend to overestimate the energy transfers from the diatomic
rotation to the inter-fragment translational degree of freedom
\cite{plarregaray:02b}.  Physically, this is due to the bending mode being
unrealistically hindered, \emph{i.e.} insufficiently damped, in the course of
fragmentation.

The previous reaction rate calculations on this PES \cite{sjklippenstein:96},
being an observable of `global' character, would very likely not depend as
strongly as the distributions we intent to reproduce.  Nevertheless, only the
construction of a new global PES or \emph{ab initio} dynamics calculations could
give a reliable answer to this issue.  Both these are beyond the scope of the
present work.

On the other hand, several modifications can be attempted in order to bring the
rotation-to-translation transfers closer to the empirical observations, under
various physical grounds.  One possible direction is to slightly change the
force constants related with the molecular modes directly responsible for such
transfers.  The alternative followed here was to rewrite the PES as
\begin{eqnarray}
\label{eq:Vlambda}
 V_\lambda(\boldsymbol{S})=xV(\boldsymbol{S})+(1-x)V^\infty(\boldsymbol{S}),
\end{eqnarray}
where $x=\exp{\left[-\lambda(S_4-R^\ddagger)\right]}$ and
$V^\infty=V_{\mathrm{CH}_2}+V_\mathrm{CO}$ is simply the asymptotic potential
interaction.  Eq.~(\ref{eq:Vlambda}) conserves both the PES at the TS and the
correct asymptotic behaviour.  In particular, $\lambda=0$ corresponds to the
original PES.  Physically sensible values for the damping parameter $\lambda$,
from calculations on triatomic systems, are usually between 2 and 3.

Trajectories were propagated on the modified PES of Eq.~(\ref{eq:Vlambda}) and
the theoretical curves corresponding to $\lambda=0.75$ and 1.5 are included in
Fig.~\ref{fig:PEtrans_cnv}.  From these, it is clear that reducing the energy
transfer between CH$_2$ rotation and inter-fragment recoil translation
invariably leads to the overall experimental trend.  The change is more strongly
pronounced at higher CO rotational excitations and all $\lambda=1.5$ curves are
in very good agreement with the experiment.

\subsection{`Inserting' the CH$_2$ rotational structure}
\label{sec2:addrot}
From Fig.~\ref{fig:PEtrans_cnv} is readily seen that the main qualitative
difference between ours (\emph{e.g.}, $\lambda=1.5$) and the experimental
results is the rotational structure observed in the latter, which relates to
that of CH$_2$($\tilde{a}^1A_1$).  In this section we show how these features
can also be theoretically reproduced by the adequate modification of the
$P^C(E_\mathrm{t};E,J,j_\mathrm{CO})$ distributions.

\subsubsection{A quasi-classical formula to get the rotational resolution}
\label{sec3:PEt5atsc}
Let us first consider the simpler though instructive case of a triatomic
unimolecular dissociation ABC~$\longrightarrow$~{A + BC($v,j$)}, at fixed total
energy $E$ and angular momentum $J$.  Here, $v,j$ represent the vibrational and
rotational states of the diatomic fragment.  In particular, we are interested in
the vibrationally-resolved product translational energy distribution (at a
sufficiently large value of the inter-fragment distance $R^\infty$).

Classically, after removing all center-of-mass coordinates, this is exactly
\begin{equation}
\label{eq:PEt3at}
 P^C(E_\mathrm{t};E,J,v)=
  \int_{\mathbf{\Gamma}}{\prod^{5}_{i=1}\Delta_i\;
   \chi(\mathbf{\Gamma})}\mathrm{d}\mathbf{\Gamma},
\end{equation}
being
\begin{equation}
 \mathbf{\Gamma}=\{R,q,\alpha,\beta,\alpha_l,\alpha_j,P,x,J',J_z,l,j\}
\end{equation}
the physically relevant set of angle-action variables (complemented with the
dissociation coordinate $R$ and its conjugate momentum $P$,
\emph{e.g.}~Eq.~(\ref{eq:gamma5at})), and
\begin{eqnarray}
 \Delta_1&=&\delta(\mathcal{H}_\mathrm{t}-E_\mathrm{t}), \\
 \Delta_2&=&\delta(\mathcal{H}-E), \\
 \Delta_3&=&\delta(J'-J), \\
 \Delta_4&=&\delta(x-v), \\
 \Delta_5&=&\frac{P}{\mu}\Theta(P)\delta(R-R^\infty),
\end{eqnarray}
where $\Delta_1$ to $\Delta_4$ explicitly account for all fixed magnitudes
while $\Delta_5$ limits integration to the \emph{outgoing flux} through the
$R=R^\infty$ surface.  As usual, $\delta$ and $\Theta$ are respectively Dirac's
$\delta$- and Heaviside functions.  In addition, the total Hamilton function has
been partitioned as
$\mathcal{H}=\mathcal{H}_\mathrm{t}+\mathcal{H}_\mathrm{rv}$, \emph{i.e.}~into
its translational and rovibrational components, with
\begin{equation}
 \mathcal{H}_\mathrm{t}=\frac{1}{2\mu}\left(P^2+\frac{l^2}{R^2}\right)+
  V_\mathrm{A,BC}.
\end{equation}
By definition, the interaction potential of A with respect to BC,
$V_\mathrm{A,BC}$, tends to zero as $R\rightarrow R^\infty$.

At last, $\chi(\mathbf{\Gamma})\mathrm{d}\mathbf{\Gamma}$ is the probability for
the system mechanical state to be in the volume $\mathbf{\Gamma}$ of phase
space.  The $\chi(\mathbf{\Gamma})$ function can be rewritten as
\begin{equation}
 \chi(\mathbf{\Gamma})=C_{JJ_zlj}\widetilde{\chi}(\mathbf{\Gamma}),
\end{equation}
where $C_{JJ_zlj}=\triangle(J,l,j)\Theta(J-|J_z|)$ holds all vector constraints,
$\triangle$ representing a triangular inequality.  $\widetilde{\chi}$ is
effectively independent of $R^\infty$ (provided this is large enough) and
carries \emph{all} the information on the full dissociation process: that coming
from both the initial state distribution and possible \emph{dynamical effects},
\emph{i.e.}~the state-to-state transition probabilities.

On recalling some basic $\delta$-function properties, partial integration in
Eq.~(\ref{eq:PEt3at}) can be shown to yield (we simplify notation by discarding
the explicit parametric dependence with respect to $E$ and $J$ in what follows)
\begin{eqnarray}
\label{eq:PEt3atv1}
 P^C(E_\mathrm{t};v)&\propto&
 \int C_{JJ_zlj}\delta\left(E-E_\mathrm{t}-E_\mathrm{rv}\right)
  \nonumber\\ && \times\overline{\chi}(E_\mathrm{t},J_z,l,j;v)\,
  \mathrm{d}J_z\mathrm{d}l\mathrm{d}j,
\end{eqnarray}
where $E_\mathrm{rv}$ is the rovibrational energy of BC and $\overline{\chi}$
is the angle/phase average of $\widetilde{\chi}$.  The centrifugal orbital term
has been neglected for it depends on $R^{-2}$ which becomes negligible as
$R\rightarrow R^\infty$.  Using the mean value theorem, this can be expressed as
\begin{equation}
\label{eq:PEt3atv2}
 P^C(E_\mathrm{t};v)=\Xi^C(E_\mathrm{t};v)P^C_{LB}(E_\mathrm{t};v),
\end{equation}
being
\begin{equation}
\label{eq:PCLB}
 P^C_{LB}(E_\mathrm{t};v)=\int C_{JJ_zlj}\delta(E-E_\mathrm{t}-E_\mathrm{rv})
 \mathrm{d}J_z\mathrm{d}l\mathrm{d}j
\end{equation}
the (classical) \emph{least-biased} vibrationally-resolved product translational
energy distribution.

Eq.~(\ref{eq:PEt3atv2}) may even look deceptively simple, though its derivation
here proves useful to disentangle the physical origin of its factors.  In a
sense, the underlying ideas can be related to the theory of surprisal analysis
of Levine and Bernstein \cite{rdlevine:87}, if the least-biased plays the role
of the prior distribution.  Eq.~(\ref{eq:PEt3atv2}) basically states that $P^C$
differs by a scaling factor (which in turn depends on $E_\mathrm{t}$ and $v$)
with respect to the least-biased value $P^C_{LB}$.  The factor itself, $\Xi^C$,
`inherits' from $\widetilde{\chi}$ all the dynamical information on the
particular process, but it is additionally `weighted' by the density of
\emph{constraint} rotational states.

In obtaining a quasi-classical analogue to Eq.~(\ref{eq:PEt3atv2}) two
approximations are made.  First, we replace the integrals in Eq.~(\ref{eq:PCLB})
by the corresponding quantum sums and second, we \emph{postulate} that if QCT
founding ideas are valid for the process under study, using $\Xi^C$ in the place
of its quantum analogue $\Xi^Q$ must be a fairly good approximation.  We should
note that the latter is not only a dynamical statement, but also a subtle
assumption on the equivalence between the classical and convoluted-quantum
density of rotational states, which is expected to hold under quite general
conditions\footnote{This is not generally true for \emph{vibrational} states,
hence we deal with vibrationally-resolved distributions.}.  By means of these
approximations one can write
\begin{equation}
\label{eq:PEt3atv3}
 P^{QC}(E_\mathrm{t};v)=\Xi^C(E_\mathrm{t};v)P^Q_{LB}(E_\mathrm{t};v),
\end{equation}
which explicitly reads
\begin{equation}
\label{eq:PEt3atsc}
 P^{QC}(E_\mathrm{t})=\sum_{\forall v,J_z,l,j}
  \frac{P^C(E_\mathrm{t};v)}{P^C_{LB}(E_\mathrm{t};v)}C_{JJ_zlj}
  \delta(E-E_\mathrm{t}-E_\mathrm{rv}),
\end{equation}
after summing over all possible final vibrational states.

Although our derivation has been made for a triatomic system, it is easy to
write the equivalent of Eq.~(\ref{eq:PEt3atsc}) for the correlated translational
energy distribution in the fragmentation of ketene
\begin{eqnarray}
\label{eq:PEt5atsc}
 P^{QC}(E_\mathrm{t};j_2)&=&\sum_{\forall v_i,J_z,l,k,j_1,\tau}
 \frac{P^C(E_\mathrm{t};v_i,j_2)}{P^C_{LB}(E_\mathrm{t};v_i,j_2)}
  C_{JJ_zlkj_1j_2\tau} \nonumber\\
 && \times\delta(E-E_\mathrm{t}-E_\mathrm{rv}),
\end{eqnarray}
where
\begin{eqnarray}
 C_{JJ_zlkj_1j_2\tau}&=&\triangle(J,l,k)\triangle(k,j_1,j_2)\Theta(J-|J_z|)
 \nonumber\\ &&\times\Theta(j_1-|\tau|).
\end{eqnarray}
that finally allows to insert the CH$_2$ rotational structure to our
vibrationally-resolved distributions in Fig.~\ref{fig:PEtrans_cnv}.  Here $j_1$
and $\tau$ constitute mere \emph{labels} for the CH$_2$ rotational energy
levels, as will become apparent in Section~\ref{sec3:erotch2}.

The ECCPST proves again particularly convenient as the $P^C_{LB}$ distribution
in Eq.~(\ref{eq:PEt5atsc}) can be directly obtained from the ensemble of
microcanonical initial conditions.

\subsubsection{CH$_2$($\tilde{a}^1A_1$) asymmetric rotor energy levels}
\label{sec3:erotch2}
Methylene has different its three principal moments of inertia
(\emph{i.e.}~$I_{aa}<I_{bb}<I_{cc}$), being thus an asymmetric rotor (AR).
There is no general quantum formula which determines the rotational energy
levels for such molecules in terms of the set of (quantum numbers) action
variables, $j_1,\kappa_1$, used for computing the corresponding initial
conditions \cite{mleo:09a}.

However, it can be proved that the rigid-rotor Hamiltonian
$\hat{\mathcal{H}}_\mathrm{rot}$ becomes block-diagonal in the matrix
representation of symmetric rotor (SR) eigenfunctions
\cite{rnzare:88,prbunker:05}.  It is also possible to show that for a given
angular momentum $J$, every AR state correlates with one prolate ($J,K_a$) and
one oblate ($J,K_c$) SR level only.  It is thus a very common practice to
\emph{label} the AR levels as $J_{K_aK_c}$, the indexes taking the values
$K_a\in\{0,1,1,2,2,\ldots,J,J\}$ and $K_c\in\{J,J,\ldots,2,2,1,1,0\}$ with
increasing energy of the terms.  As it is easily seen, one could alternatively
use a single label defined by $\tau=K_a-K_c$, which in turn spans the $2J+1$
values from $-J$ to $J$.  $\tau$ rather than $\kappa_1$ has been used in
Eq.~(\ref{eq:PEt5atsc}) to stress the fact that this label has no direct
correspondence to the $\kappa_1$ action.

In our calculations, spectroscopic CH$_2$($\tilde{a}^1A_1$) energy levels were
used when available and complemented with theoretical AR calculations at higher
energies \cite{hpetek:87a,ubley:93}.  The error in theoretical values could be
of several inverse centimetres, especially at large values of $j_1$ and $\tau$.

\begin{figure}[!t]
\includegraphics[width=85mm]{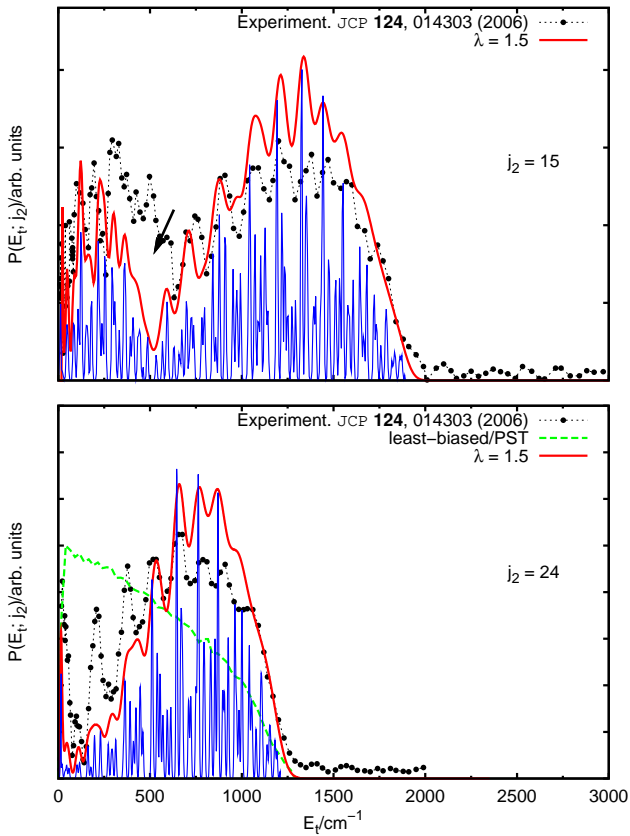}
\caption{(Colour online) Correlated product translational energy distributions
 after excitation with a 308~nm laser: QCT results with rovibrational
 resolution.  Comparison with the experiment.}
\label{fig:PEtrans}
\end{figure}

\section{DISCUSSION}
\label{sec:discussion}
Rotational structures are barely visible in the measurements for $j_2=4$ and 9,
being strongly blurred due to the contribution of adjacent rotational states.
Fig.~\ref{fig:PEtrans} then shows the application of Eq.~(\ref{eq:PEt5atsc}) to
$\lambda=1.5$ theoretical distributions, corresponding to $j_2=15$ and 24, as a
test on the capabilities of the method.  Within each panel, two theoretical
curves correspond to convolutions with different strengths.  One of them
(darker, red) has been chosen to closely match the experimental resolution
\cite{avkomissarov:06} while the other (blue) shows the typical fine structures
which are attainable with our method.

The empirical rotational structure is fairly well reproduced by the QCT
calculations, being all theoretical and experimental peaks correcly aligned.
The larger disagreement is observed within the upper panel, in the region shown
by an arrow.  This area corresponds to the matching between the low energy tail
of the $x_2=0$ state and the high energy component of $x_2=1$, being the former
the main contribution to it.  The discrepancy between our curve and the
experiment is thus a secondary effect due to the overestimation of
rotation-to-translation transfers already discussed in
Section~\ref{sec3:lambdaPES}.  It cannot be seen in Fig.~\ref{fig:PEtrans_cnv}
as the convolution used there artificially `populates' the missing rotational
states.

The least-biased distribution corresponding to $j_2=24$ is also depicted in
Fig.~\ref{fig:PEtrans}, obtained from the whole set of initial conditions.  This
distribution is directly comparable with the PST calculations shown in Fig.~8 of
Ref.~\cite{avkomissarov:06}.  In fact, application of Eq.~(\ref{eq:PEt5atsc}) in
order to `add' a rotational structure to this curve would trivially reduce it to
the quantum version of the least-biased/PST distribution.  At last, by comparing
this with the final theoretical curves, it is readily seen how the suppression
observed in the low-$E_\mathrm{t}$ part are correctly described by QCT.

It is important to address one last point referring to the insertion of the
rotational structures in Section~\ref{sec2:addrot}.  In principle, one might
think that a different, simpler approach could be used: To additionally fix all
rotational quantum numbers and propagate separate ensembles for each
semiclassical state, very much as we have done here for each scissor excitation
or can be found elsewhere, \emph{e.g.}~\cite{fjaoiz:07a}.  However, with respect
to the initial conditions there is a basic difference in the way the vibrational
and rotational modes are determined \cite{mleo:09a}.  Using the normal-mode
approximation, the former are assigned precise \emph{energies} corresponding to
the quantum values.  This means that normal mode displacements and conjugate
momenta are calculated to exactly match the desired energy in the particular
mode.  On the other hand, classically, after fixing the \emph{angular momentum
vector} for a particular fragment, there remains an indetermination in the
rotational energy due to the vibration-rotation coupling.  In other words, there
are different inertia tensors due to dissimilar normal-mode displacements which
will result in different energies even for the same angular momentum.  In the
case of ketene, for example, calculations after additionally fixing the values
of $j_1$ and $\kappa_1$ yield broad structures in the correlated translational
energy distributions which are several hundred inverse centimetres wide.  This
would completely blur the rotational structures observed in
Fig.~\ref{fig:PEtrans}.  The indetermination should be smaller for symmetric
tops---where a formula exists that determines the energy in terms of the angular
momentum and one of its components---but still would, in general, strongly limit
the QCT rotational resolution.

\section{SUMMARY AND CONCLUSIONS}
\label{sec:summary}
A QCT algorithm has been presented that allows to simulate indirect polyatomic
processes involving exit-channel effects up to rovibrational resolution without
the use of any binning or weighting procedure.  It consists of three stages: (1)
initial conditions are started in terms of angle-action variables to
individually select the experimental states to be studied, which are later
transformed into Cartesian coordinates for trajectory integration; (2)
trajectories are propagated using an inverse dynamics simulation in the spirit
of the exit-channel corrected phase space theory of Hamilton and Brumer; and (3)
finally, an approximate quasi-classical formula is used to accurately
incorporate possible rotational structures into the quasi-classical product
distributions.  Of these, only the second step is linked to the statistical
assumptions inherent to the study of indirect processes, being the other two
of general applicability.

The method was introduced and its capabilities illustrated with the simulation
of highly-resolved correlated translational energy distributions from recent
experiments in the photo-dissociation of ketene at 308~nm
\cite{avkomissarov:06}.  After slightly modifying the PES to deal with an
overestimation of the energy transfer among rotational and translational degrees
of freedom, calculated distributions with both vibrational and rotational
resolution are in good agreement with the experiment.

\section*{ACKNOWLEDGEMENTS}
Support from an Inter-University Agreement on International Joint Doctorate
Supervision between the Instituto Superior de Tecnolog\'{\i}as y Ciencias
Aplicadas, Cuba and the Universit\'e Bordeaux 1, France, as well as the PNCB/2/4
project of the Departamento de F\'{\i}sica General in the Cuban institution, are
gratefully acknowledged.

\providecommand*{\mcitethebibliography}{\thebibliography}
\csname @ifundefined\endcsname{endmcitethebibliography}
{\let\endmcitethebibliography\endthebibliography}{}

\end{document}